\documentclass[usenatbib]{mnras}
\usepackage[utf8]{inputenc}
\usepackage{natbib,epsfig,cite,amsmath,amssymb,graphicx}
\usepackage{array, xcolor, caption, multirow}

\def\e10{\eta_{10}}

\def\iso#1#2{\mbox{${}^{#2}{\rm #1}$}}

\def\b1#1{\iso{B}{1#1}}
\def\msun{\mbox{$M_\odot$}}

\def\beq{\begin{equation}}
\def\eeq{\end{equation}}
\def\beqar{\begin{eqnarray}}
\def\eeqar{\end{eqnarray}}
\def\simlt{\lower.5ex\hbox{$\; \buildrel < \over \sim \;$}}
\def\simgt{\lower.5ex\hbox{$\; \buildrel > \over \sim \;$}}
\def\simpropto{\lower.2ex\hbox{$\; \buildrel \propto \over \sim \;$}}

\title[Cosmological evolution of Nitrogen]{Cosmological evolution of the Nitrogen abundance}
 
\author[E. Vangioni et al.]
{Elisabeth Vangioni$^{1}$\thanks{e-mail:vangioni@iap.fr},
Irina Dvorkin$^	{1,6}$,
Keith A. Olive$^{2}$, 
Yohan Dubois$^{1}$,
\newauthor Paolo Molaro$^{3}$, Patrick Petitjean$^{1}$, Joe Silk$^{1, 4}$ and Taysun Kimm$^{5}$\\
$^{1}$Sorbonne Universit\'es, UPMC Univ Paris 6 et CNRS, UMR 7095, Institut d'Astrophysique de Paris, 98 bis bd Arago, 75014 Paris, France\\
$^{2}$William I. Fine Theoretical Physics Institute, University of Minnesota, Minneapolis, MN 55455, USA\\
$^{3}$INAF – Osservatorio Astronomico di Trieste, Via G.B. Tiepolo 11, I-34143 Trieste, Italy\\
$^{4}$Department of Physics and Astronomy, The Johns Hopkins University, Baltimore MD21218 USA\\
$^{5}$Department of Astronomy, Yonsei University, 50 Yonsei-ro, Seodaemun-gu, Seoul 03722, Republic of Korea\\
$^{6}$Max Planck Institute for Gravitational Physics (Albert Einstein Institute), Am M\"uhlenberg 1, Postdam-golm, 14476 Germany
}
 
\begin{document}

\pagerange{\pageref{firstpage}--\pageref{lastpage}} \pubyear{2017}
\maketitle
\label{firstpage}

\begin{abstract}
The abundance of nitrogen in the interstellar medium is a powerful probe of star formation processes over cosmological timescales.  Since nitrogen can be produced both in massive and intermediate-mass stars with metallicity-dependent yields, its evolution is challenging to model, as evidenced by the differences between theoretical predictions and observations. In this work we attempt to identify the sources of these discrepancies using a cosmic evolution model. To further complicate matters, there is considerable dispersion in the abundances from observations of damped Ly$\alpha$ absorbers (DLAs) at $z\sim 2-3$. 
We study the evolution of nitrogen with a detailed cosmic chemical evolution model and find good agreement with these observations, including the relative abundances of (N/O) and (N/Si). We find that the principal contribution of nitrogen comes from intermediate mass stars, with the exception of systems with the lowest N/H, where nitrogen production might possibly be dominated by massive stars. This last result could be strengthened if stellar rotation which is important at low metallicity can produce significant amounts of nitrogen.
Moreover, these systems likely reside in host galaxies with stellar masses below $10^{8.5}M_\odot$. We also study the origin of the observed dispersion in nitrogen abundances using the cosmological hydrodynamical simulations Horizon-AGN. 
We conclude that this dispersion can originate from two effects: difference in the masses of the DLA host galaxies, and difference in the their position inside the galaxy.

\end{abstract}


\section{Introduction}
Measurements of chemical abundances in the interstellar medium (ISM) are a powerful probe of galaxy evolution and star formation processes. The total metallicity content, dominated by oxygen, reflects the star formation history, as well as the history of gas accretion and galactic outflows. To understand cosmic chemical evolution, theoretical models have been developed 
\citep[e.g.][]{daigne06,2012MNRAS.421...98D,2013ApJ...772..119L,2015ApJ...808..129L,2016MNRAS.455.1218B} and studies of metal abundances in different galactic environments have unveiled important physical processes, such as galactic outflows \citep[e.g.][]{2016MNRAS.455.1218B} and the origin of star formation quenching \citep{2015Natur.521..192P}. These are based on 
galactic chemical evolution models (\citet{1972A&A....20..383T,1975MNRAS.172...13P,1978ApJ...221..554T,2004oee..symp...85M,2006ApJ...653.1145K}) that now follow the cosmological evolution of the abundances
as a function of redshift.

Observations of individual element abundances, as well as their relative abundances are particularly informative, as they can constrain the nucleosynthetic processes as well as specific mass ranges of stars reponsible for their production. Nitrogen is a particularly interesting (and challenging) element in this regard since it can be produced in both massive and intermediate-mass stars, which release it into the ISM on different timescales. Thus, the study of the nitrogen abundance in the ISM, and in particular its relative abundance with respect to oxygen, which is produced mostly by massive, short-lived stars, poses interesting challenges for chemical evolution models.

Nitrogen can be produced as a \emph{primary} element, in a sequence of nuclear reactions that involve only hydrogen and helium present in the star. In this case its abundance grows in proportion to that of oxygen and (N/O) remains constant as (O/H) grows \citep{Talbot74}, assuming metallicity-independent yields. Additionally, \emph{secondary} nitrogen is produced from CNO elements present in the star, so that the nitrogen yield is metallicity-dependent and (N/O) is no longer constant  \citep{Clayton83, Arnett96}. Interestingly, the observed (N/O) abundance exhibits a low-metallicity plateau, which is present in damped Ly$\alpha$ absorbers (DLAs) \citep{centurion03,2003astro.ph..1407M} and is now evident in extragalactic HII regions with SDSS data \citep[see][and references therein]{2016MNRAS.458.3466V}, which suggest primary nitrogen production.  However observations of the nitrogen abundance in  DLAs \citep{1995ApJ...451..100P,1996A&A...308....1M,Lu98,pettini02, centurion03, petitjean08, cooke11, zafar14} have revealed a significant dispersion in (N/H) at any given redshift
which adds another layer of difficulty.
Thus there is some uncertainty in the main production sites of nitrogen (massive or intermediate-mass stars), and it is still unclear which of these sites, if any, produces primary nitrogen in significant amounts. 
 Indeed in \citet{centurion03}  it was  concluded that DLAs do show evidence of primary N production at low metallicities and in addition that there are two plateaus.
The [N/$\alpha$] ratios 
are distributed in two groups: about 75\% of the DLAs show {a mean value of} [N/$\alpha$] {= --0.87} 
and  about 25\%  shows ratios at 
[N/$\alpha$] {= --1.45}.
The lower group may originate from massive stars due to the 
tight linear correlation between N/H and $\alpha$ elements. The group with higher [N/$\alpha$], provides evidence for primary production of intermediate mass  stars.
The transition between 
the   low   and  
high-N  DLAs   could result from  the different lifetimes of massive and intermediate mass stars.

 The nitrogen abundance in the ISM was extensively studied in the context of galaxy evolution models \citep[e.g.][]{1993A&A...277...42P,Fields98,1999A&A...346..428P,Henry00,Tissera02,2003A&A...397..487P,2003MNRAS.339...63C,2005A&A...437..429C,Molla06,Gavilan06,Wu13,2016MNRAS.458.3466V}.
It should be noted that there is an inherent uncertainty in chemical evolution models as stemming from the uncertainty in 
stellar nitrogen abundances. Nevertheless, 
 these studies confirmed that while the origin of the nitrogen abundance is a mix of primary and secondary sources, the total nitrogen budget is dominated by intermediate-mass stars in star-forming galaxies (as well as in other systems including DLAs)
  except perhaps at low metallicites.
  We note that various theoretical models typically utilize different sets of chemical yields, making the comparison of results problematic.

The goal of the current study is to examine the uncertainties in different stellar evolution models and their effect on the predicted cosmic evolution of the nitrogen abundance, specifically regarding DLAs observations. We use a uniform cosmic chemical evolution framework. We also 
compare the results of our semi-analytic model with the output of cosmological hydrodynamical simulations.

The paper is organized as follows: the data used for model comparison is described in Section~\ref{sec:data} and we review the production of nitrogen and oxygen in several stellar evolution models in Section~\ref{sec:nucleosynthesis}. In particular, we discuss the dependence of nitrogen and oxygen yields on stellar mass, metallicity, and rotation velocity. In Section~\ref{sec:meanevolution}  we show how these yields are implemented in our cosmic chemical evolution model and we describe the Horizon-AGN simulations in Section~\ref{sec:HAGN}. Our results for the mean evolution of nitrogen and oxygen abundances in the interstellar medium as a function of redshift are given in Section~\ref{results}. In Section~\ref{sec:dispersion}, we discuss the dispersion in nitrogen abundances in DLAs  in  hydrodynamical cosmological Horizon-AGN simulations~\citep{duboisetal14} of galaxies of different masses and at different redshifts.
We conclude and discuss future prospects in Section~\ref{sec:conclusions}.

\section{Data}
\label{sec:data}

 Nitrogen has been observed in various galactic environments and at different redshifts. Here we summarise the observations used in the current study (see for example  Fig.~\ref{fig:NO_log}).

Nitrogen is accurately measured in the HII regions of dwarf irregular   and blue compact dwarf galaxies. We collect here the observations of \citet{vanZee98a,vanZee06} and of \citet{Izotov04,James15,Berg12,2012A&A...546A.122I} respectively. 
The flat slope in the (N/O) vs. (O/H) plane found in  star-forming dwarf galaxies  below 12+ log(O/H) = 7.7   led to the conclusion that primary nitrogen has to be produced in massive stars, that promptly release it into the ISM \citep[][]{1995ApJ...445..108T,Berg12,2016MNRAS.458.3466V}. However, at low metallicity, intermediate mass stars are also expected to 
produce nitrogen and as we will see, our detailed stellar evolution models do not generically predict sufficient primary nitrogen from massive stars to match the bulk of the observations.

Nitrogen can be also accurately measured in the DLAs which probe lower metallicities  and high redshifts. We use here the 
compilation of high redshift data coming from  \citet{zafar14}  and references therein.  The nitrogen measurement compilation includes 
a sample of 108 systems but actual measurements  of  N and O, S, Si abundances are  for  27 DLAs and sub-DLAs, of which 18 critically  drawn from the literature.  The typical error bar of [N/$\alpha$] in DLAs is  of order 0.02 dex. This extended sample shows the [N/$\alpha$] bimodal behaviour suggested in previous studies (see  \citet{centurion03}). 
 To note that 
 the high [N/$\alpha$] plateau in the DLAs is consistent with the low metallicity tail of HII regions of dwarf irregular  and blue compact dwarf galaxies  but it extends  to lower metallicities. Indeed,  it is remarkable that observations in these very different environments and different redshifts establish a continuous relation in the N-$\alpha$ plane (although there is significant scatter; see Fig.~\ref{fig:NO_log}).
On the other hand, 
the low [N/$\alpha$] plateau  (the lowest ever observed in any astrophysical site) gathered in \citet{zafar14} suggest the presence of a floor in [N/$\alpha$] abundances, which may indicate a primary nitrogen production from fast rotating, massive stars in young or unevolved systems.

Unfortunately N measurements in galactic halo stars are quite uncertain. 

 Historically first measurements of N in metal poor stars were based on the NH 3360 band and provided almost solar abundances. For instance, [N/Fe] were found -0.5 and -0.2 in the  well known halo stars HD 140283 and HD 64090, a result which was interpreted
 as a primary origin of N, and as a product of massive stars  \citep{1984ApJ...279..220T,1986MNRAS.221..911M}.

 \citet{2004A&A...416.1117C} and \citet{2005A&A...430..655S}  provided  measurements based on the  CN 3880 and NH 3360 band respectively.    \citet{2005A&A...430..655S} used the C/N abundance to separate the stars with mixed products from the CN cycle. There is a systematic  offset  for the abundances based on the two bands with the NH-abundances being of 0.4 dex higher.  When NH abundances are rescaled to match CN abundances  the   unmixed stars   with [Fe/H] $<$ -3.4 show [N/O] $\approx$-0.9  overlapping the  DLA's values  though with significant scatter (cfr Fig 17 in \citet{2005A&A...430..655S}).    \citet{2005A&A...430..655S}  note that   accurate values of {\it gf}  and dissociation energy for NH as well as studies of 3D model atmosphere effects on the band strength are needed to assess the accuracy of the N abundances and therefore the stellar  determinations will not be used in this paper.

Other observations  of N abundances in nearby galaxies such as M33 and NGC 55 \citep{2007A&A...470..843M,2017MNRAS.464..739M} have abundances consistent with those used here. 
Recent surveys such as SDSS DR12 using 100000 star forming galaxies \citep{2016ApJ...828...18M} or SDSS IV MaNGA using 550 nearby galaxies \citep{2017MNRAS.469..151B} show an interesting correlation between the N/O ratio and the stellar mass of these galaxies which could be more fundamental than the relation N/O vs O/H at a redshift around 0.
Finally note that N abundances have been measured in planetary nebulae and these studies can be used to test 
AGB stellar models \citep{2006ApJ...651..898S,2010ApJ...714.1096S,2017MNRAS.468..272C,2017MNRAS.471.4648V}  at different metallicities/masses.

{\bf }

\section{Stellar nucleosynthesis of nitrogen}
\label{sec:nucleosynthesis}

While there are many uncertainties regarding nitrogen yields in stars of different mass, the nuclear reactions leading to its formation are fairly well understood. 
Nitrogen is mainly produced in the CN branch of the CNO cycle \citep{Clayton83, Arnett96} via the following chain of reactions: $^{12}C (p, \gamma) ^{13} N(\beta+, \nu) ^{13} C (p, \gamma) ^{14} N$ (note that nitrogen can also be produced in the ON cycle by transformation of $^{16}O$, but at a much
slower rate). The reaction $^{14} N(p,\gamma) ^{15} O$ which depletes nitrogen has a relatively low cross section enabling $^{14}N$ to accumulate with time. In this formation scenario nitrogen is a secondary element, whose production rate depends on the abundance of the CNO elements initially present in the star.

Alternatively, nitrogen can be produced as a primary element \citep{Talbot74}. In this case the reactions are the same as above, but the sequence of events is different: first, some $^{12}C$ is synthesized by the $3\alpha$ reaction in a helium burning region, and then this new $^{12}C$ is transported to the hydrogen burning region, where the CNO cycle converts it to nitrogen. Thus, primary nitrogen is likely to be formed in stars with a helium burning core and a CNO burning shell, provided there is some mechanism that transports carbon between the two regions. 

Observationally, the nucleosynthetic origin of nitrogen can be deduced by comparing its abundance to that of other elements. The abundance of primary nitrogen is proportional to that of the other primary elements (assuming metallicity-independent yields), while if nitrogen is produced as a secondary element, the increase in its abundance is proportional to the initial CNO content. Overall, the abundance of secondary nitrogen is proportional to the square of the CNO content. Note, however, that the evolution of nitrogen abundance in the ISM of a given galaxy is more complex, as it depends on the chemical evolution history of the galaxy and the inflow of primordial gas that dilutes the ISM.

Furthermore, various stellar evolution models predict different nitrogen yields, in particular, primary N yields from massive stars depend on Z and as a result might not produce a plateau. That is, if the yield of N/O ia a function of metallicity, the resulting evolution of N/O vs O/H may not be constant and a plateau 
is not reproduced, thus confusing the interpretation of primary vs secondary origin of N. This adds an additional complication to the analysis.

The yields have been tested in a variety of chemical evolution models \citep{chiappini06,Molla06,2010A&A...522A..32R,2011MNRAS.414.3231K}. These studies focused on reproducing the Milky Way chemical abundances which are used to calibrate different models.
We next summarize several sets of nucleosynthetic yields used in this work for comparison.

\subsection{Massive stars}
\label{sec:massivestars}

In order to estimate the theoretical uncertainties in nitrogen yields in massive stars we consider three different stellar evolution models: \citet[][; WW95]{ww95}, \citet{Chieffi04} and \citet{Nomoto06}. Specifically, WW95 present the evolution of massive stars at $5$ different metallicites ($Z/Z_{\odot}=0, 10^{-4}, 10^{-2}, 0.1, 1$) and masses from $11M_{\odot}$ to $40M_\odot$. \citet{Chieffi04} produce another set of explosive yields for masses in the range $13 - 35 M_\odot$ at $6$ different metallicities ($Z= 0,10^{-6},10^{-4}, 10^{-3}, 6\times 10^{-3}, 0.02$). \citet{Nomoto06} also study the mass range $13-35M_\odot$ for $4$ different metallicities: $Z= 0, 10^{-3}, 0.004, 0.02$. For illlustration, table \ref{table1} compares typical yields produced by these models at three metallicities ($Z/Z_{\odot}=0, 0.001, 1$) and two masses: $15M_{\odot},30M_{\odot}$ by interpolating the published yields. 
For higher masses, we extrapolate the published yields. As we will show below, given the power-law initial mass function (IMF) used in our models, our results are not sensitive to these extrapolated yields.
There are significant differences between the models, especially at low metallicities. These discrepancies are the result of the different models for the microphysics of pre-supernova evolution (such as the treatment of the convective layers) and different reaction rates, for example $^{12}C(\alpha,\gamma)^{16}O$ \citep[see the discussion in][]{Chieffi04}. Another major difference is the production of primary nitrogen, which is not included in WW95. Additional yields can be found in \citet{2010ApJ...724..341H}, \citet{2012ApJS..199...38L}, and  \citet{2017arXiv170601913L}. The uncertainties in stellar yield assumptions were
also studied in \citet{2010A&A...522A..32R,2015MNRAS.451.3693M,2016ApJ...824...82C,2016MNRAS.463.3755C,2017ApJ...835..224A}.

\subsection{Intermediate-mass stars}
\label{sec:imstars}

We consider two stellar evolution models that target the evolution of intermediate-mass stars: \citet{hg97} and \citet{Karakas10}. \citet{hg97} present the evolution of stars at $5$ different metallicites ($Z = 0.001, 0.004, 0.008, 0.02, 0.04$; note that the last value is super-solar) and masses in the range $0.8- 8M_\odot$. \citet{Karakas10} present another set of yields for masses in the range $1-6M_\odot$ at $4$ different metallicities ($Z= 0.0001, 0.004, 0.008, 0.02$). Note that a recent study \citep{Fishlock14} has presented yields at $Z = 0.001$ for the same mass range.  Table \ref{table2} presents typical yields for three metallicities ($Z=0.0001, 0.004, 0.02$) and two masses, $2M_{\odot}$ and $7M_\odot$. The results of these two models are similar (note that the largest differences are again at the low-metallicity end). As we will show in what follows, intermediate-mass stars play a dominant role in producing the nitrogen observed in the interstellar medium, therefore we do not expect large uncertainties in the modeled total nitrogen abundance. Note however that, since the yields are metallicity-dependent, uncertainties in the total metallicity (mainly due to oxygen abundance) may affect also the result for nitrogen.

 Note that iron and silicon yields coming from SNIa are included in our model (respectively $0.5M_\odot$ $0.1M_\odot$). Details can be found in \citet{daigne06}.

\begin{table*}
\caption{Typical nitrogen and oxygen yields of intermediate-mass stars as a function of stellar mass and initial metallicity in different stellar evolution models (masses are in solar mass units): VdH97: \citet{hg97}, Karakas10: \citet{Karakas10}.}
\begin{center}
\begin{tabular}{c|c|c|c|c|c|c|c|c|c|c|c|c}
Metallicity & \multicolumn{2}{c|}{$10^{-4}Z_{\odot}$} & \multicolumn{2}{c|}{$10^{-4}Z_{\odot}$} & \multicolumn{2}{c|}{$4\cdot 10^{-3}Z_{\odot}$} & \multicolumn{2}{c|}{$4\cdot 10^{-3}Z_{\odot}$} & \multicolumn{2}{c|}{$Z_{\odot}$} & \multicolumn{2}{c}{$Z_{\odot}$}\\
\hline
Stellar Mass & \multicolumn{2}{c|}{2} & \multicolumn{2}{c|}{7} & \multicolumn{2}{c|}{2} & \multicolumn{2}{c|}{7} & \multicolumn{2}{c|}{2} & \multicolumn{2}{c}{7}\\
\hline 
 Element & N & O & N & O & N & O & N & O & N & O & N & O\\
\hline
VdH97  & $10^{-4}$ & $1.5\cdot 10^{-3}$ & $0.024$ & $0.0036$ & $0.001$ & $0.001$ & $0.062$ & $0.0005$ & $0.0045$ & $0.015$ & $0.096$ & $0.06$ \\
Karakas10 & $7.8\cdot 10^{-5}$ & $5.8\cdot 10^{-4}$ & $0.039$ & $0.001$ & $0.000685$ & $0.00293$ & $0.063$ & $0.0044$ & $0.003$ & $0.013$ & $0.049$ & $0.047$ \\
\hline
\label{table2}
\end{tabular}
\end{center}
\end{table*}

\subsection{Stellar rotation}
\label{sec:rotation}

Stellar rotation, particularly important at low metallicities, can strongly affect the nucleosynthetic yields \citep{mm02a, mm02b, meynet04}. In particular, \citet{mm02a} found that rotating, low-metallicity ($Z = 10^{-5}$) stars naturally produce primary nitrogen as a consequence of enhanced mixing between the hydrogen-burning shell and the core. Table \ref{table3} compares the yields calculated by \citet{mm02b} of rotating ($v=300$ km/s) and non-rotating stars of different masses ($2-20M_\odot$) at $Z = 10^{-5}$. It can be seen that both nitrogen and oxygen are produced in larger amounts in rotating stars, the difference with the non-rotating case reaching two orders of magnitude. Note however that this finding cannot have a significant impact on the evolution of the total nitrogen abundance, most of which is taking place in higher-metallicity environments. We will nevertheless explore the impact of this set of yields, were it to hold for the entire range of metallicities, below.

\begin{table*}
\caption{Typical nitrogen and oxygen yields of massive stars as a function of stellar mass and initial metallicity in different stellar evolution models (masses are in solar mass units):  WW95: \citet{ww95}, Chieffi04: \citet{Chieffi04}, Nomoto06: \citet{Nomoto06}. Note that the models strongly disagree at $Z=0$.}
\begin{center}
\begin{tabular}{c|c|c|c|c|c|c|c|c|c|c|c|c}
 Metallicity  & \multicolumn{2}{c|}{0} & \multicolumn{2}{c|}{0} & \multicolumn{2}{c|}{$10^{-3}Z_{\odot}$} & \multicolumn{2}{c|}{$10^{-3}Z_{\odot}$} & \multicolumn{2}{c|}{$Z_{\odot}$} & \multicolumn{2}{c}{$Z_{\odot}$} \\
\hline
 Stellar Mass & \multicolumn{2}{c|}{15} & \multicolumn{2}{c|}{30} & \multicolumn{2}{c|}{15} & \multicolumn{2}{c|}{30} & \multicolumn{2}{c|}{15} & \multicolumn{2}{c}{30}\\
\hline 
 Element & N & O & N & O & N & O & N & O & N & O & N & O \\
\hline 
WW95  & $3.63\cdot 10^{-6}$ &  0.4 & $3.53\cdot 10^{-3}$ &  4.35 & $4.85\cdot 10^{-3}$ & 0.555 & $1.08\cdot 10^{-2}$ & 4.42 & $0.054$ & 0.68 & $0.104$ & 4.88\\
Chieffi04 & $2.95\cdot 10^{-7}$ & 0.346 & $6.95\cdot 10^{-2}$ & 3.04 & $2.99\cdot 10^{-3}$ & 0.527 & $5.63\cdot 10^{-3}$ & 3.75 & $0.051$ & 0.516 & $0.0857$ & 3.89\\
Nomoto06 & $1.86\cdot 10^{-3}$ & 0.773 & $1.64\cdot 10^{-6}$ & 4.81 & $3.58\cdot 10^{-3}$ & 0.294 & $6.19\cdot 10^{-3}$ & 5.33 & $0.062$ & 0.16 & $0.102$ & 3.22\\
\hline
\label{table1}
\end{tabular}
\end{center}
\end{table*}

\begin{table*}
\caption{Typical nitrogen and oxygen yields (in solar mass units) of rotating and non-rotating stars for different stellar masses \citep{mm02b}. In all the models the metallicity taken to be $Z = 10^{-5}$ and the velocity of rotation $v = 300$ km/s.}
\begin{center}
\begin{tabular}{c|c|c|c|c|c|c|c|c}\\
 Stellar Mass & \multicolumn{2}{c|}{2} & \multicolumn{2}{c|}{7} & \multicolumn{2}{c|}{15} & \multicolumn{2}{c}{20} \\
\hline 
 Element & N & O & N & O & N & O & N & O\\
\hline
$v=0$ &$1.7\cdot 10^{-6}$ & $ 1.9\cdot 10^{-5}$ & $1.6\cdot 10^{-5}$ & $ 1.2\cdot 10^{-3}$ & $3.5\cdot 10^{-5}$ & $0.19$ & $4.6\cdot 10^{-5}$ & $0.56$\\
$v=300$ km/s & $6.4\cdot 10^{-4}$ & $1.1\cdot 10^{-3}$ & $4.2\cdot 10^{-3}$ & $5.6\cdot 10^{-3}$ & $4\cdot 10^{-4}$ & $0.39$ & $3.4\cdot 10^{-4}$ & $0.99$\\
\label{table3}
\end{tabular}
\end{center}
\end{table*}

\section{Models}

\subsection{Cosmic chemical evolution model}
\label{sec:meanevolution}

In order to describe the cosmic chemical evolution of the interstellar medium we use the model developed by \citet{daigne04, daigne06}, \citet{2009MNRAS.398.1782R} and \citet{vangioni15} as outlined in the following. The initial gas content of galaxies is taken to be equal to the cosmic mean
$f_{baryon}=\Omega_b/\Omega_m$, where $\Omega_b$ and $\Omega_m$ are the densities of baryons and total dark matter, respectively, in units of the critical density of the Universe. The gas is assumed to be primordial and metal-free and the calculation begins at a redshift $z = 20$. The model then follows two gas reservoirs, the intergalactic matter (IGM) and the ISM. Matter flows from IGM to ISM as the galaxies form, where baryons are assumed to follow dark matter. 
In  \citet{daigne06}, the mean baryon accretion rate in each region was taken to be proportional to the fraction of 
baryons in structures, $f_{coll}$, and can be expressed as 
\begin{equation}
a_\mathrm{b}(t)  =  \Omega_\mathrm{b}\left(\frac{3H_{0}^{2}}{8\pi G}\right)\ \left(\frac{dt}{dz}\right)^{-1}\ \left|\frac{d f_{coll}}{dz}\right|
\end{equation}
where $f_{coll}(z)$ is given by the hierarchical model of structure formation \citep{ps74}, 
\begin{equation}
f_{coll}(z) = \frac{\int_{M_\mathrm{min}}^{\infty} dM\ M f_\mathrm{PS}(M,z)}{\int_{0}^{\infty} dM\ M f_\mathrm{PS}(M,z)}\ .
\label{eq:fb}
\end{equation}
and we assume that the minimum mass of dark matter haloes for star-forming structures is $10^{7}~M_\odot$. 
Accreted matter is assumed to be primordial and metal-free.
Once inside galaxies, baryons form stars with rate $\psi(t)$ which we calibrate to observations, as described below. We assume a Salpeter IMF, $\Phi(m)$, with slope $x = 1.35$, for $m_{inf}\leq m \leq m_{sup}$ with $m_{inf}=0.1 M_{\odot}$ and $m_{sup}=100M_{\odot}$. 
Baryons can flow from structures back to the IGM due to galactic winds or feedback from supernovae (SNe). To sum up, the evolution of the total baryonic mass in the IGM and ISM is given by 
\begin{equation}
\frac{dM_{IGM}}{dt}=-a_b(t)+o(t)
\end{equation}
and 
\begin{equation}
\frac{dM_{ISM}}{dt}=-\psi(t)+e(t)+a_b(t)-o(t),
\end{equation}
where $\psi(t)$ is the cosmic star formation rate (SFR), $e(t)$ is the rate at which stellar mass is returned to the ISM by mass loss or stellar deaths and $o(t)$ is the baryon outflow rate from structures into the IGM, which is obtained by energy arguments assuming the 
conservation of a fraction of the energy released in SNII and assuming outflows with escape velocities. 
The outflow rate is actually the sum of two separate outflows \citep{daigne04}, 
$o(t)=o_\mathrm{w}(t)+o_\mathrm{sn}(t)$. The first one,
$o_\mathrm{w}(t)$, is a global outflow powered by the stellar
explosions (galactic wind) and is similar to that described in
\citet{1997ApJ...476..521S}. 
The
 second term, $o_\mathrm{sn}(t)$ corresponds to the fraction $\alpha$
 of stellar supernova ejecta which is flushed directly out of the
 structures, resulting in metal-enhanced winds as first proposed by
 \citet{1986ApJ...305..669V}.  Note also, $o_\mathrm{w}(t)$ carries the chemical composition of the ISM, 
whereas $o_\mathrm{sn}(t)$ has  the chemical composition of the supernovae.
The outflow rate is
always small compared to the baryon accretion rate and detailed expressions for $o(t)$ are given in  \citet{daigne06}.

In addition, we follow the chemical composition of the ISM and the IGM as a function of time as described in
\citet{daigne04}. In particular, we do not use the instantaneous recycling approximation but calculate the rate at which gas is returned to the ISM
including the effect of stellar lifetimes and computing stellar yields for each element and for different stellar mass ranges.
The lifetimes of intermediate mass stars ($0.9<M/\msun <8$) are taken
 from \citet{mm89} and  from
\citet{schaerer02} for more massive stars. 
Further details on the chemical evolution model can be found in \citet{daigne04,daigne06} and \citet{vangioni15}.

\begin{figure}
\begin{center}
\epsfig{file=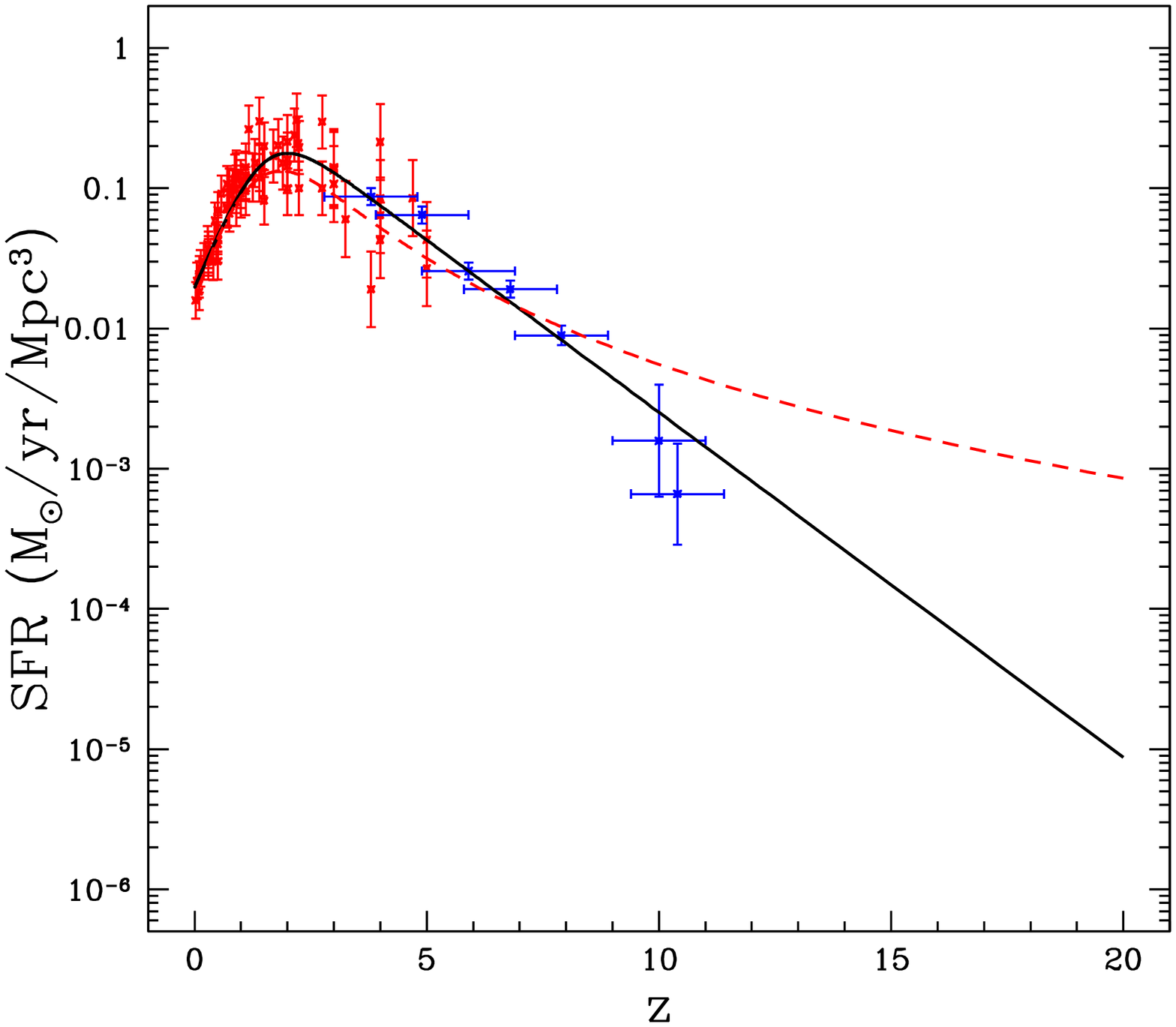, height=3in}
\vskip -1 cm
\epsfig{file=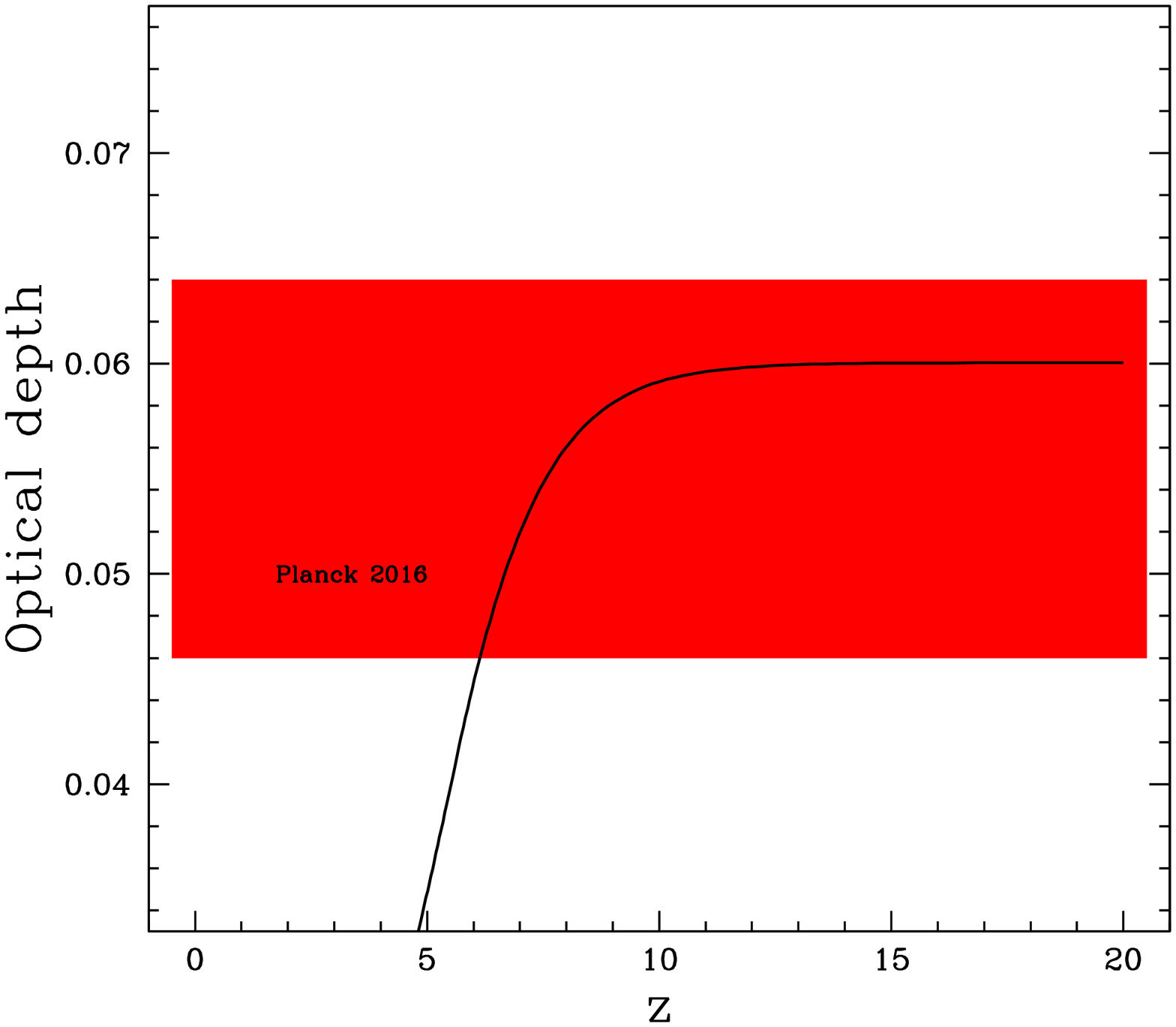, height=3in}
\end{center}
\caption{\emph{Upper panel:} The SFR as a function of redshift used in our model (black line) compared to observations: \citet{Behroozi13} (red points),
\citet{Bouwens14, O14} (blue points). The SFR in \citet{Madau14} is shown for comparison (red dashed line). \emph{Lower panel:} Evolution of the optical depth to reionization as a function of redshift. The observational constraint is indicated by a red horizontal strip \citep{Planck161, Planck162}.}
\label{fig:sfr1}
\end{figure}

In typical models of galactic chemical evolution, the SFR resides among the set of physical quantities, with assumed forms
(e.g., a decaying exponential in time or proportional to the gas density) with parameters which are fit by the observations of 
the chemical abundances. Flexibility in the SFR was also adopted in the early cosmic chemical evolution models in \citet{daigne04,daigne06} where SFR parameters were also required to fit the observed SFR at high redshift. However, 
over the last decade there have been significant improvements in the observations of the SFR out to high redshift, $z \sim 10$ and 
in fact, there is very little flexibility in the SFR out to $z \sim 3$. In  \citet{2009MNRAS.398.1782R}, a fixed SFR was used
based on the \citet{springel03} form 
\begin{equation}
\psi(z) = \nu\frac{a\exp(b\,(z-z_m))}{a-b+b\exp(a\,(z-z_m))}\,.
\end{equation}
with parameters fit directly to the observed SFR. 
Here, we use the cosmic SFR modeled in \citet{vangioni15} with
$\nu = 0.178 $  M$_{\odot}$/yr/Mpc$^{3}$, 
$z_m = 2.00, a = 2.37$, and $b = 1.80$. Fig.~\ref{fig:sfr1} (upper panel) shows the SFR in our model (black line) and observations compiled by \citet{Behroozi13}, as well as high-redshift measurements by \citet{Bouwens14, O14}. For comparison, we also show the SFR from \citet{Madau14}.

The SFR can also be constrained by using the optical depth to reionization, which depends on the rate of production of ionizing photons by massive stars. We calculate the optical depth to reionization $\tau$ as described in \citet{vangioni15}, in particular we use the tables in \citet{schaerer02} for the number of photons produced by massive stars and assume an escape fraction of $f_{esc}=0.2$. The resulting optical depth is shown in Fig.~\ref{fig:sfr1} (bottom panel) and compared to the constraints obtained from measurements of the cosmic microwave background \citep{Planck161, Planck162}. Note that the SFR of  \citet{Madau14},
is flatter than the SFR used in our model and produces slightly higher rates than observed at $z \sim 10$.
Consequently, the optical depth is also larger than that observed by Planck, unless $f_{esc}$ is as low as 0.05. Note that there are other parameters which introduce additional uncertainties in the model prediction of the optical depth.
Finally, we stress that this model does not follow galaxy properties individually but averaged properties of the ISM and of the IGM within a sub-volume of the Universe.  As a result, metallicities generally remain subsolar.
For further details on the evolution of the metallicity with redshift in the context of a merger tree model, see: \citet{dvorkin15,2016MNRAS.458L.104D}. Note that in this model, we do not follow the chemical evolution of individual galaxies, but the averages over relatively large volumes, treating each region of the Universe as a closed box. We are currently working on an updated model where each galaxy is treated individually, as part of a larger project that also studies other chemical species, and will present this model in a forthcoming publication.

\subsection{Horizon-AGN simulation}
\label{sec:HAGN}

The details of the simulation can be found in~\citet{duboisetal14} and references therein, but we recall here the main aspects of the simulation. The Horizon-AGN simulation is a hydrodynamical cosmological simulation of a flat Lambda Cold Dark Matter universe performed with the adaptive mesh refinement code {\sc ramses}~\citep{teyssier02} using a WMAP-7-like cosmology~\citep{komatsuetal11} with $\Omega_{\rm m}=0.272$, $\Omega_{\Lambda}=0.728$, $\sigma_8=0.81$, $\Omega_{\rm b}=0.045$, $H_0=70.4\,\rm km\,s^{-1}\,Mpc^{-1}$, and $n_{\rm s}=0.967$.
The box size is $100\, \rm h^{-1}\, Mpc$ filled $1024^3$ dark matter particles with a mass resolution of $8\times 10^7 \, \rm M_\odot$. The mesh refinement is triggered according to a quasi-Lagragian criterion when the mass within a cell is 8 times the dark matter mass resolution down to a spatial resolution of $1 \,\rm kpc$.
The Horizon-AGN simulation includes a metal-dependent gas cooling with a ultra-violet background heating after reionization at $z=10$, a Schmidt star formation law of constant efficiency of 2 per cent, feedback from stars including stellar winds, type II and type Ia supernovae assuming a Salpeter-like initial mass function \citep{kaviraj17}, and feedback from active galactic nuclei together with the self-consistent growth of black holes following an Eddington-limited Bondi-Hoyle-Littleton accretion rate.

The mass loss from stellar winds and SNe is modelled using {\sc starburst99} \citep{vazquez05}.
Specifically, Horizon-AGN adopts the `Evolution’ model \citep{leitherer92}, which computes the outflow rates from stellar winds with the Padova tracks plus thermally pulsing AGB stars \citep{bertelli94,girardi00}. This model is based on the SN yields from \citet{ww95} for massive stars with mass $8\le M/M_\odot\le 40$. For low-mass to intermediate-mass ($0.1\le M/M_\odot\le7$) stars, the chemical yields are taken from the chemical abundance at the stellar surface computed from the Padova stellar tracks \citep{girardi00}. These yields are similar to those coming from \citet{Karakas10} and \citet{hg97}.
Note that most ($\sim 80\%$) of the nitrogen is produced before the age of 50 Myr via SNe, whereas intermediate mass stars account for only $20\%$, at that time. The yields for SN Type Ia are taken from \citet{Nomoto06}, but their contributions to the total production of nitrogen are negligible.

To extract the metal mass content of the galaxies, we include all gas cells within the effective radius of the galaxy and with gas density above $0.1 \,\rm H\, cm^{-3}$, which is our gas density threshold for star formation.

\section{Results}

\subsection{Evolution of nitrogen abundance with redshift}
\label{results}

\begin{figure}
\centering
\epsfig{file=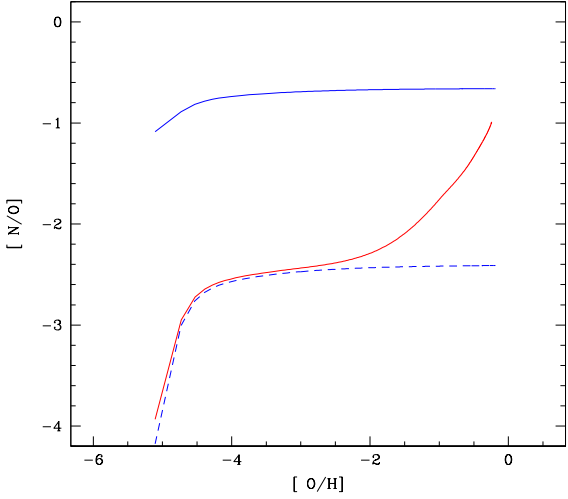, height=3in}
\caption{The evolution of nitrogen in the  ISM (total) produced in massive stars in the semi-analytical model. Blue lines: corresponding to the constant yields from \citet{Nomoto06} for $Z = Z_{\odot}$ and $Z = 0$ for solid and dashed lines, respectively. Red line: metallicity-dependent yields from \citet{Nomoto06}. Note that in the latter case [N/O] increases sharply with metallicity, as can be anticipated from Table \ref{table1}.}
\label{fig:NOtestSNII}
\end{figure}

In order to study the production of nitrogen and its abundance in the ISM we implement the stellar evolution models discussed in Section \ref{sec:nucleosynthesis} in our galaxy evolution model. We use the yields discussed in Sections \ref{sec:massivestars} and \ref{sec:imstars} for stars in the mass ranges $8 < M/M_\odot < 40$ and $0.9 < M/M_\odot < 8.0$, respectively. An interpolation is made between different metallicities and masses, and tabulated values are extrapolated for masses above $40\,M_{\odot}$, which correspond to a small fraction of the population of massive stars (i.e. $\left(40^{-x}-100^{-x}\right)/\left(8^{-x}-100^{-x}\right)\simeq 8\%$). We adopt the solar abundances from \citet{asplund09}. We stress the importance of including metallicity-dependent yields, rather than constant yields, as demonstrated in Fig. \ref{fig:NOtestSNII}. In this figure, we compare the resulting ratio of 
[N/O], assuming constant yields (blue curves) to a metallicity dependent yield (red curve).  In each case, 
we employ the yields from \citet{Nomoto06}. For constant yields, we show results for both solar (solid curve)
and metal-free (dashed curve) yields.  As one might expect from Table \ref{table1}, the produced N/O ratio is significantly larger 
in solar metallicity stars than in metal-free stars. The model with metal-dependent yields  interpolates between the two.

\begin{figure}
\begin{center}
\epsfig{file=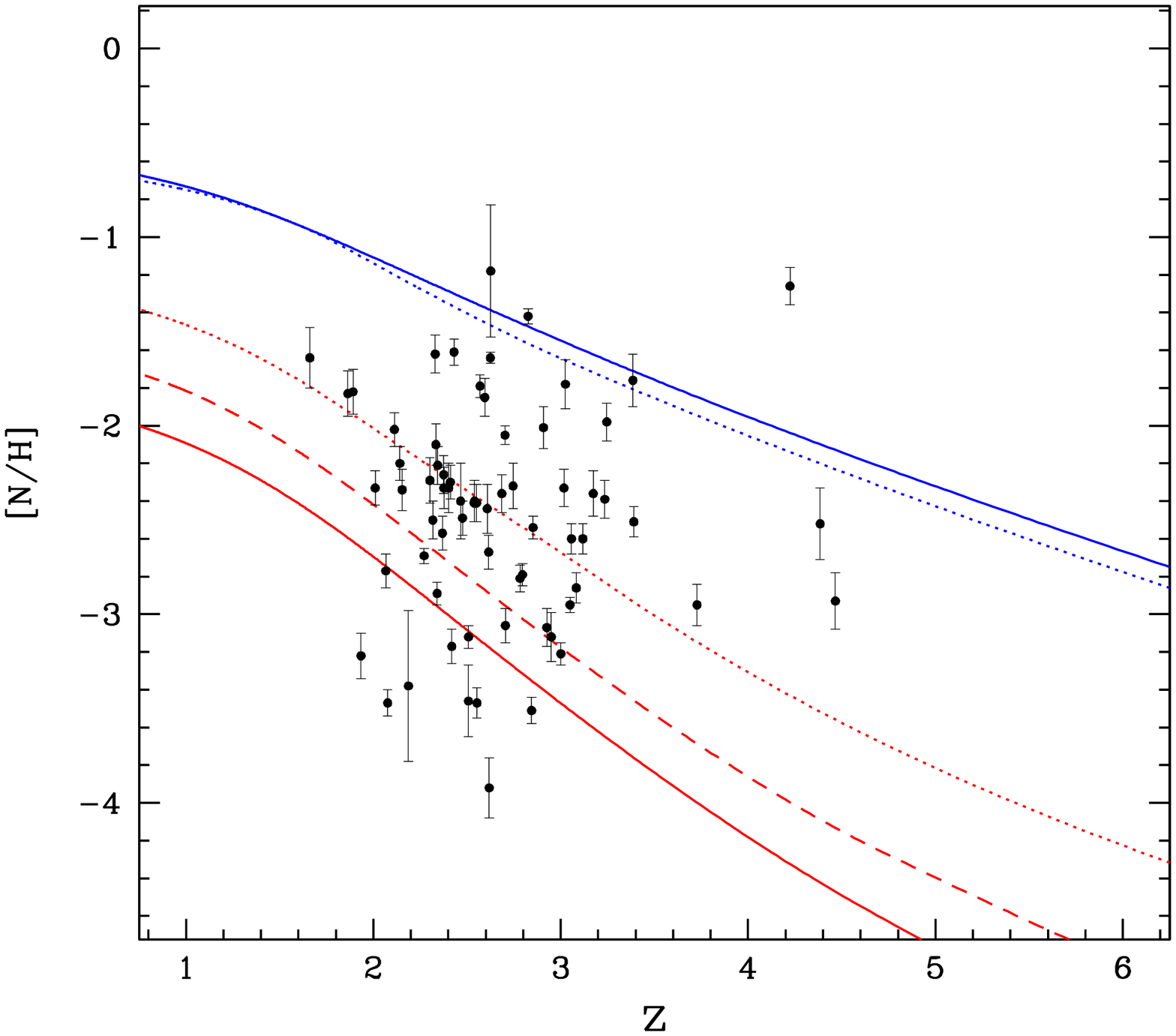, height=3in}
\vskip -1 cm
\epsfig{file=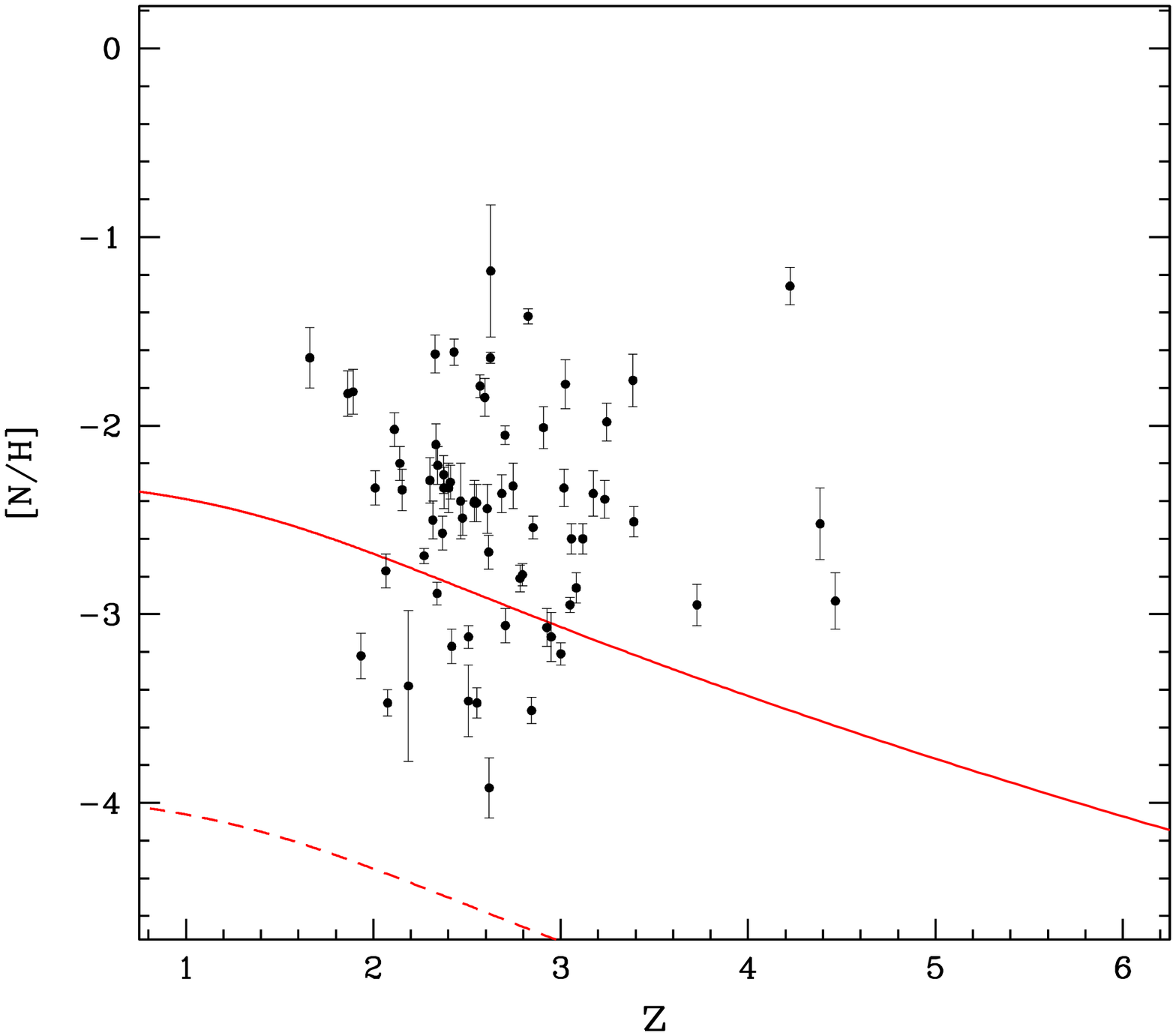, height=3in}
\end{center}
\caption{Evolution of nitrogen abundance with redshift for different stellar evolution models. \emph{Upper panel:} Nitrogen produced in massive stars only, with yields taken from \citet{Chieffi04}, \citet{ww95} and \citet{Nomoto06} (red solid, dashed and dotted lines, respectively) and the total nitrogen budget (blue lines). The yields for intermediate-mass stars are taken from \citet{Karakas10} and \citet{hg97} for the dotted and solid blue lines, respectively, the yields for massive stars are taken from \citet{Nomoto06} in both cases. 
\emph{Lower panel:} The evolution of nitrogen abundance assuming all stars are rotating (solid) or non-rotating (dashed) and using constant yields, taken as the yields from \citet{mm02b} at $Z = 10^{-5}$.
Data taken from \citet{zafar14} and references therein (black points).}
\label{fig:Nz1}
\end{figure}

We now compare the results of our model to the observed abundance measurements, focusing on the comparison between different stellar evolution models and the contribution of massive vs. intermediate-mass stars to the nitrogen content of the ISM. 
Fig.~\ref{fig:Nz1} (upper panel) shows the evolution of nitrogen abundance in our model with different sets of yields: blue dotted and solid lines correspond to models that include both intermediate mass stars (IMS) with yields taken from \citet{Karakas10}) and \citet{hg97}, respectively,  and massive stars (MS), with yields taken from \citet{Nomoto06}. As can be anticipated from Table \ref{table1}, the two sets of yields for intermediate-mass stars produce very similar results.  
Red solid, dashed and dotted lines correspond to models that include only the contribution of massive stars, with the yields taken from \citet{Chieffi04}, \citet{ww95} and \citet{Nomoto06} respectively. In this case, the differences in nitrogen and oxygen yields shown in Table \ref{table1} are translated into a relatively large vertical offset between the red curves. It is clear, however, that intermediate-mass stars dominate the production of nitrogen at all redshifts and this contribution is needed to explain
the abundances with the highest values of [N/H].  Note that the differencies in the yields of massive stars do not affect the uncertainty in the total nitrogen budget. This result confirms the finding of \citet{Wu13} for massive star-forming galaxies. We also note that the scatter in [N/H] observed at any given redshift is much greater than the observational uncertainty. This scatter can originate from differences in structure formation histories of the different galaxies that host these DLAs, as we discuss below, including different accretion and star formation histories. In addition, the scatter, for now, masks any trend in the observations
of [N/H] with redshift as predicted by the models.

\begin{figure}
\begin{center}
\epsfig{file=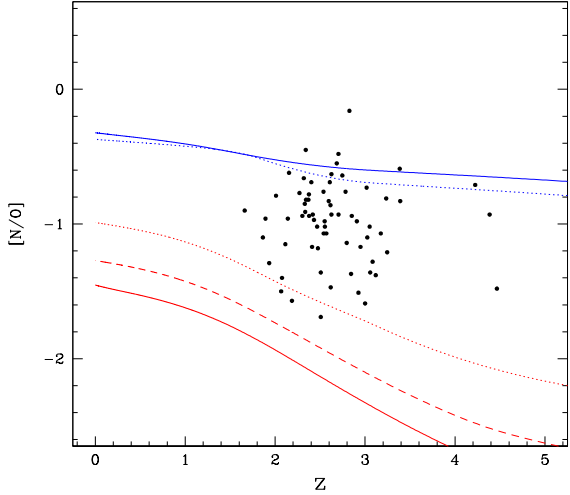, height=3in}
\end{center}
\caption{As in  Fig.~\ref{fig:Nz1}, the relative abundance [N/O] as a function of redshift. }
\label{fig:NOz}
\end{figure}

As in the upper panel of Fig.~\ref{fig:Nz1}, Fig.~\ref{fig:NOz} shows the evolution of the relative [N/O]  abundance in our model as a function of $z$, with different sets of yields. We note that massive stars (red curves, modulo uncertainties related to stellar yields) ) which form the first nitrogen atoms can reproduce the lowest part of DLA observations ([N/O] = -1.5). It is interesting to note that blue and red curves frame the  bulk of observations possibly corresponding to a progressive increase in nitrogen coming from intermediate mass stars.

The lower panel in Fig.~\ref{fig:Nz1} shows the effect of stellar rotation, where we used \emph{constant} yields (computed for $Z = 10^{-5}$) taken from \citet{mm02b}. While, as evident also from Table \ref{table3}, rotating low-metallicity stars produce significant amounts of primary nitrogen, this model is in fact not realistic since the actual metallicity at the relevant redshifts is much higher than $10^{-5}$. Further studies of rotating stars including yield tables at higher metallicity are required in order to elucidate their role in the cosmic nitrogen enrichment.

\begin{figure}
\begin{center}
\epsfig{file=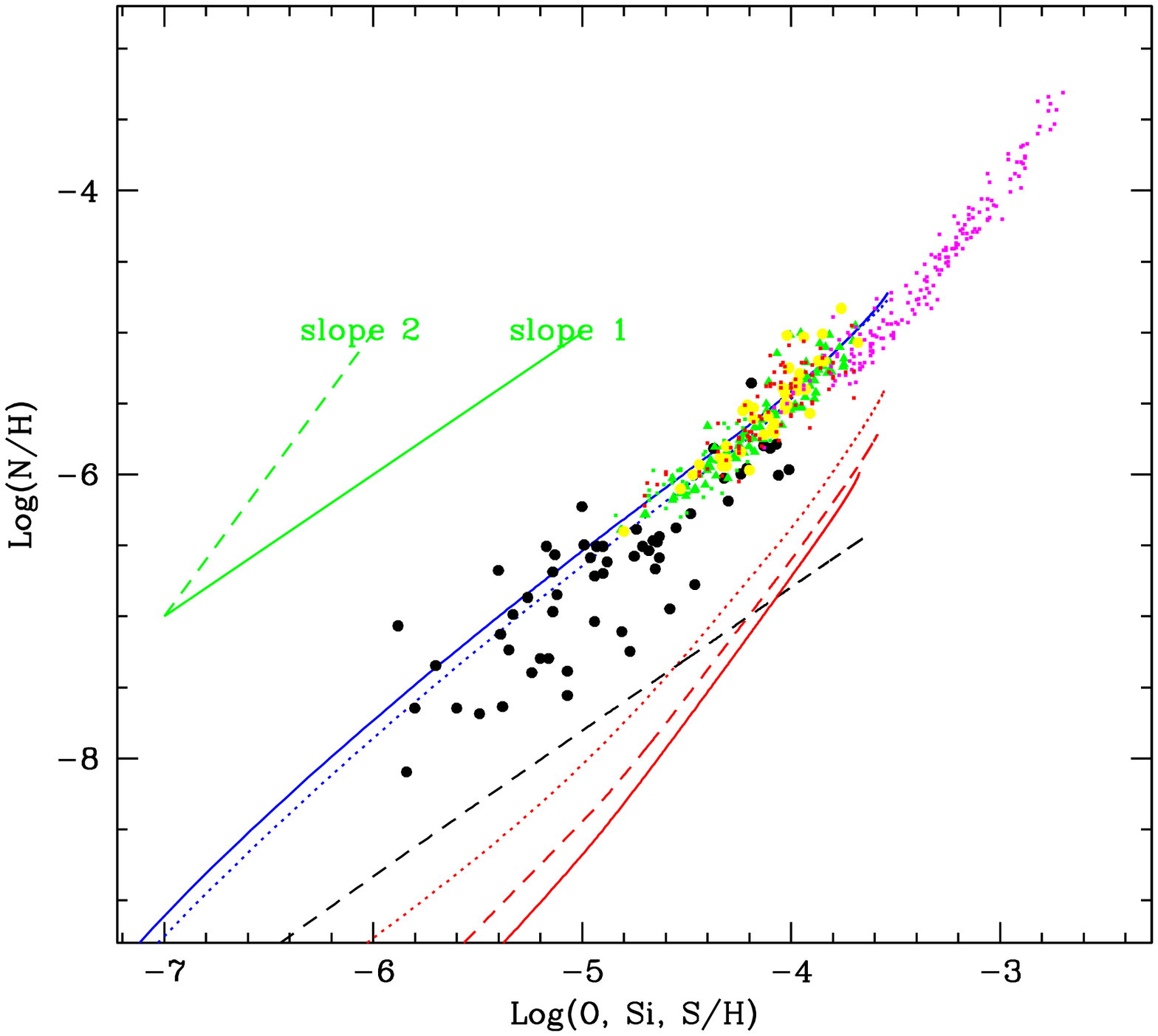, height=3in}
\end{center}
\caption{The nitrogen abundance as a function of metallicity in our model with different sets of yields (line notation as in Fig. \ref{fig:Nz1}) compared to observations in DLAs \citet{zafar14} (black points), HII regions of spiral galaxies \citep[magenta points][]{vanZee98a,vanZee06} and blue compact dwarf galaxies \citep[green, blue and yellow points, respectively;][]{Izotov04,James15,Berg12,2012A&A...546A.122I}. Dashed black line corresponds to the case of rotating stars, with constant yields (at $Z=10^{-5}$) taken from \citet{mm02b}.}
\label{fig:NO_log}
\end{figure}

Fig. \ref{fig:NO_log} shows the nitrogen abundance as a function of metallicity in DLAs. The data have been detailed in section \ref{sec:data}. As one can see, the data are well reproduced by our model, shown by the blue lines. As shown previously, relatively little nitrogen is produced in massive stars (red lines).  We also note that when available, we plot data using oxygen abundances.  However,
in some cases, either Si or S is used as a surrogate for the metallicity. The curves, nevertheless show the evolution of N/H
vs O/H as predicted by the model. The deficiency in a uniform data set introduces additional uncertainty and scatter
when comparing with O/H.  As noted above, our abundances represent a mean value which remain sub-solar,
and thus we cannot expect to reproduce the abundances of $z=0$ HII regions of spiral galaxies shown by the magenta points, in this context.

\begin{figure}
\begin{center}
\epsfig{file=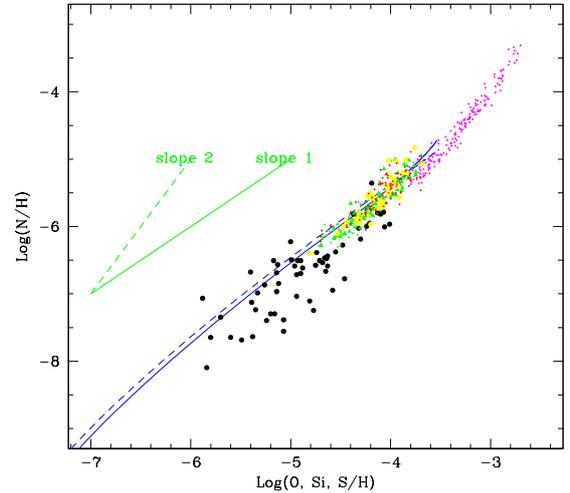, height=3in}
\end{center}
\caption{As in Fig. \ref{fig:NO_log} showing a comparison of the model using the yields of \citet{Nomoto06} and \citet{hg97}
when yields above 40 M$_\odot$ have been extrapolated (solid) and when the yields are held constant 
above 40 M$_\odot$. }
\label{fig:NOx_log}
\end{figure}

 Fig. \ref{fig:NOx_log} shows the effect of our extrapolation of stellar yields above 40 M$_\odot$. The blue solid 
curve is that shown in Fig. \ref{fig:NO_log} where the yields above 40 M$_\odot$ have been extrapolated. 
However as remarked earlier, the fraction of stars in this mass range is rather limited and we are not very sensitive to this
procedure. The blue dashed curve in Fig. \ref{fig:NOx_log} shows the calculated abundances when the yields above 
40 M$_\odot$ are held constant.  As one can see, the effect is minimal. 

It is interesting to note that our models with only massive stars (red curves)
do not exhibit the characteristic slope of 1 associated with primary production. As we have stressed earlier, 
this is due to the metallicity-dependent yields used in the calculation. In contrast, 
as seen by the black dashed curve, when metallicity-independent yields are used, the slope is indeed found to be 1.
One sees a slight steepening starting at a metallicity of roughly Log(O/H) $\sim -3.5$.

We further emphasize that due to the substantial dispersion in the data for O/H vs redshift (or more precisely
$\alpha$/H vs redshift), that a direct mapping of N/H vs $z$ to N/H vs $\alpha$/H is not possible. That is, since there is no one-to-one
relation between $\alpha/H$ and redshift, points will shift relative to model predictions when comparing results in
Figures \ref{fig:Nz1} and \ref{fig:NO_log}. 

\begin{figure}
\begin{center}
\epsfig{file=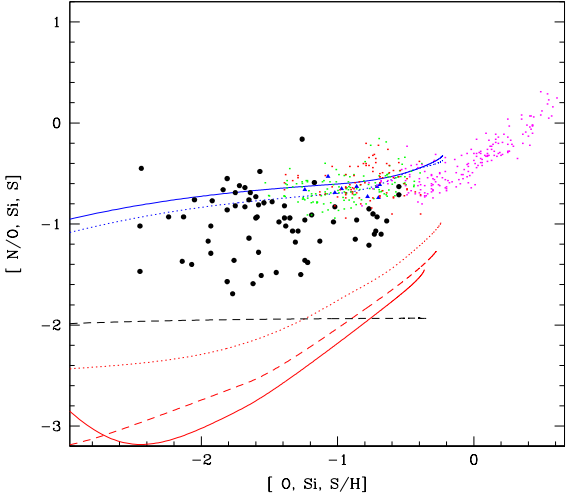, height=3in}
\end{center}
\
\caption{As in  Fig. \ref{fig:NO_log}, the relative abundance [N/O] as a function of metallicity.}
\label{fig:NO}
\end{figure}

Bearing this ambiguity in mind, the relative abundance of nitrogen and oxygen is further explored in Fig.~\ref{fig:NO} which also exhibits the scatter observed in DLAs (black points) and compact dwarf galaxies (green, blue and yellow points). It can be seen that the ratio [N/O] for nitrogen that originates in massive stars is well below the observational points. Thus, according to our model, the bulk of the nitrogen observed in DLAs and dwarf galaxies originates from intermediate-mass stars. 
A similar conclusion was reached by \citet{Henry00} who found the bulk of cosmic nitrogen to be formed in intermediate
mass stars as a primary
source at low metallicity and secondary source at higher metallicity.
We see
once again that the model with rotating stars with constant yields is flat when plotting [N/O] in contrast to the 
models with metallicity-dependent yields.  Note that 
our models do not attain solar abundances (of either O/H or N/H) as the curves represent average abundances (throughout 
the universe) and are thus heavily weighted by low mass objects with subsolar abundances. 

\begin{figure}
\begin{center}
\epsfig{file=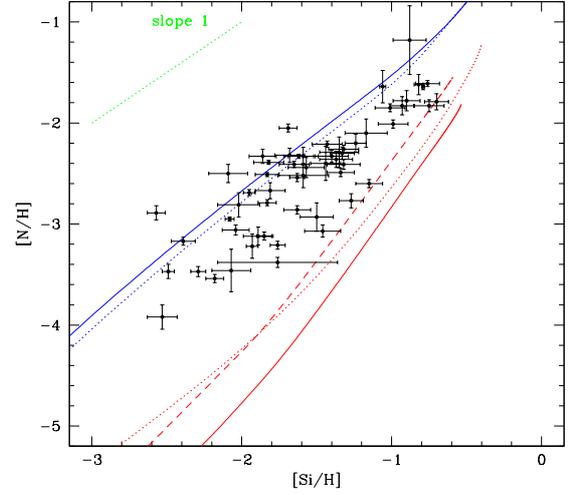, height=3in}
\vspace*{-1cm}
\epsfig{file=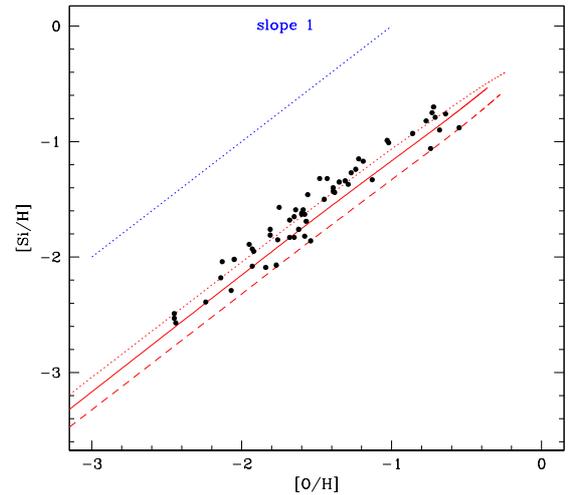, height=3in}
\end{center}
\
\caption{\emph{Upper panel:}  As in Fig. \ref{fig:Nz1} showing the evolution of [N/H] as a function of [Si/H] in our model. Data is taken from the compilation of \citet{zafar14} (black points). \emph{Lower panel:} The correlation between Si and O abundances. }
\label{fig:NSi}
\end{figure}

Finally, we use our model to explore the relative abundances of nitrogen and other elements. The upper panel in Fig.~\ref{fig:NSi} shows the tight correlation between [N/H] and [Si/H] (albeit with a considerable dispersion), reproduced by our model. Note that the observed correlation between [Si/H] and [O/H], shown in the lower panel of Fig. \ref{fig:NSi}, is much tighter, since both O and Si are produced in massive stars and are simultaneously released into the ISM.

\subsection{Dispersion in nitrogen abundances}
\label{sec:dispersion}

As can be seen in Figs. \ref{fig:Nz1} and \ref{fig:NO}, there is a significant dispersion in nitrogen abundance at any given redshift. Here we study the dispersion due to the difference in the masses of DLA host galaxies.

The masses of the DLA host galaxies may vary across our sample, and while they are very difficult to measure, they are expected to affect the DLA properties. Indeed, a mass-metallicity relation, similar to that observed in galaxies in emission, might be present in DLAs \citep{2013ApJ...769...54N}. The semi-analytic models presented above are not suitable to study this issue since they do not resolve individual galaxies. In this Section we therefore use the outcome of the Horizon-AGN simulation~\citep{duboisetal14} 
described above to estimate the dispersion in nitrogen abundance in the ISM of $z\sim 2$ galaxies.

\begin{figure}
\begin{center}
\epsfig{file=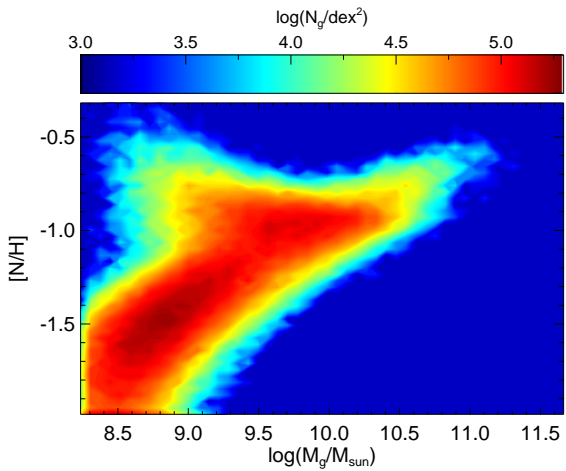, height=2.5in}
\end{center}
\caption{The nitrogen abundance in the cold gas of galaxies as a function of galaxy stellar mass in Horizon-AGN at $z = 2$. The shading corresponds to the relative number of galaxies in the sample in each abundance-mass bin.}
\label{fig:sims}
\end{figure}

The result of the abundance of nitrogen in galaxies as a function of the galaxy stellar mass in Horizon-AGN at $z=2$ is shown in Fig.~\ref{fig:sims}. We notice first that there is very little dispersion at larger masses ($M > 10^{10.5}M_\odot$) with
the amount of dispersion increasing at lower masses.  Below $10^{8.5}M_\odot$, the amount of dispersion is approximately
1.5 dex and is slightly less what is observed.  
Note that the observations of the lowest DLA abundances of N/H can be well explained if their mass corresponds to a stellar mass of $10^{8.5}M_\odot$ or less.

\begin{figure}
\begin{center}
\epsfig{file=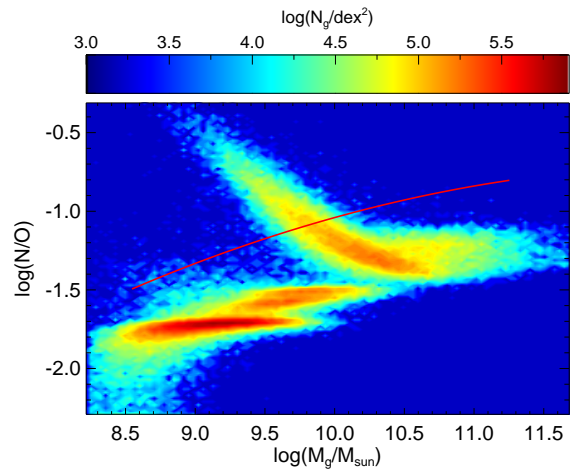, height=2.5in}
\end{center}
\caption{ The nitrogen over oxygen ratio in the cold gas of galaxies as a function of galaxy stellar mass in Horizon-AGN at $z = 0$ compared with the CALIFA observations from~\citet{perez16} (red solid line). The shading corresponds to the number density of galaxies in the sample in each abundance-mass bin.}
\label{fig:sims_no}
\end{figure}

Figure~\ref{fig:sims_no} shows the nitrogen to oxygen ratio as a function of galaxy stellar mass at $z=0$ in Horizon-AGN compared to the CALIFA data from \citet{perez16}.
The amount of nitrogen relative to oxygen is on average lower in Horizon-AGN with respect to the data at any mass. Note that we have also tested the effect of AGN feedback on the amount of nitrogen in the cold gas in galaxies using the Horizon-noAGN simulation~\citep{dubois16, peirani17} and we found little difference with the simulation including AGN feedback, a result in agreement with recent simulations from~\citet{taylor15}.

\begin{figure}
\begin{center}
\epsfig{file=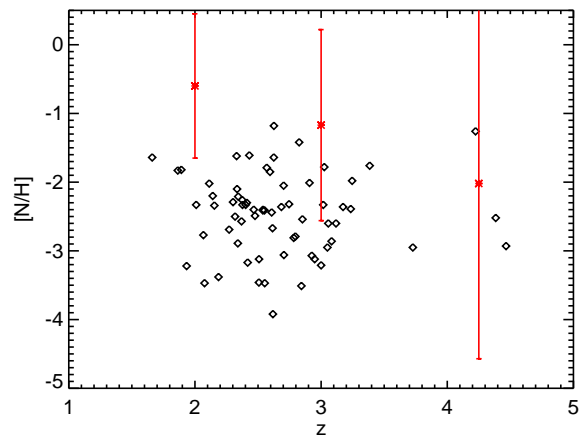, height=2.5in}
\end{center}
\caption{Nitrogen abundance as a function of redshift for data points from~\citet{zafar14} (black diamonds) and for the Horizon-AGN simulation (red symbols with 1-$\sigma$ standard deviation). The result of Horizon-AGN is obtained with the compilation of all possible line-of-sights passing through cold gas in galaxies at the three different plotted redshifts. }
\label{fig:simnhvsz}
\end{figure}

Finally, we study the effect of differences in DLA position within the host galaxy. We draw several line-of-sights through our simulation box, 
 one every $1\,\rm kpc$ in the x and y coordinates, and compute the nitrogen abundance in the cold ISM gas along each line-of-sight (at $z=2$ there are $\sim7\times 10^6$ line-of-sights containing cold gas with nitrogen).
We repeat this procedure for three different redshifts and compare the average and the standard deviation with data points from~\citet{zafar14}.
As can be seen  in Fig.~\ref{fig:simnhvsz}, the Horizon-AGN simulation predicts higher values of the nitrogen abundance at $z=2$ and $3$ compared with observations, but its value is comparable to the few data points at $z=4$.

Finally the dispersion of the nitrogen line-of-sight in Horizon-AGN is large $\sim 1 dex$ at $z=2$ and $3$ and $\sim 2 dex$ at $z=4$, which is very close to the dispersion observed in the data points for the lowest redshift range.

This is the result of having a large variety of possible line-of-sights going through galaxies of various masses, that are different stages of their chemical evolution, as well as line-of-sight probing different regions within the same galaxy: either gas with old population of stars in the center of galaxies, or freshly accreted gas from the IGM forming young stars in the outskirts.

\section{Conclusions}
\label{sec:conclusions}

In this paper we explored the evolution of the nitrogen abundance in the ISM and its dispersion using a cosmological galaxy evolution model. In particular, we explored several sets of nucleosynthetic yields currently used by various groups and showed that the bulk of nitrogen in the ISM is produced by intermediate-mass stars. At very low metallicity, nitrogen production 
can be explained by massive stars production
 while at higher metallicity nitrogen is essentially produced by IMS.
 Since the yields of this class of objects are consistent between models (contrary to the case of massive stars) it follows that the total nitrogen mass produced by a given stellar population can be calculated with reasonably high accuracy.

Furthermore, we explored the sources of dispersion in the nitrogen abundance in DLAs using the Horizon-AGN hydrodynamical simulations and concluded that it can be caused by the difference in the masses of the host galaxies. In particular, we found that  this dispersion grows with decreasing galaxy mass. Another source of dispersion is the difference in the lines of sights that corresponds to different DLAs. 
While these effects contribute to the observed dispersion, further work is needed in order to understand the relative contribution of each effect.
 In a forthcoming publication, we use an updated model where each galaxy will be treated individually (obtaining the evolution as a function of galaxy stellar mass), together with a larger project that also studies other chemical species.

\section*{Acknowledgements}

We thank warmly Georges Meynet for very fruitful discussions.
This work is made in the ILP LABEX (under reference ANR-10-LABX-63) supported by French state funds managed by the ANR within the Investissements d'Avenir programme under reference ANR-11-IDEX-0004-02. 
The work of  K.A.O. and M.P. was supported in part
by DOE grant DE--SC0011842 at the University of Minnesota. TK is supported by the National Research Foundation of Korea to the Center for Galaxy Evolution Research (No. 2017R1A5A1070354).

\bibliographystyle{mnras}
\bibliography{Nevolution}
\label{lastpage}
\end{document}